# Investigations of a THGEM-based imaging detector

**M. Cortesi**[a*]**, R. Alon**[a]**, R. Chechik**[a]**, A. Breskin**[a]**, D. Vartsky**[b]**, V. Dangendorf**[c]

[a] *Department of Particle Physics, Weizmann Institute of Science,*
  *76100 Rehovot, Israel*
[b] *Soreq NRC,*
  *81800 Yavne, Israel*
[c] *Physikalisch Technische Bundesanstalt (PTB),*
  *Bundesalle 100, 38116, Braunschweig, Germany*
  *E-mail*: marco.cortesi@weizmann.ac.il

ABSTRACT: We present the results of our recent studies on a Thick Gas Electron Multiplier (THGEM)-based imaging detector prototype. It consists of two 100x100 mm$^2$ THGEM electrodes in cascade, coupled to a resistive anode. The event location is recorded with a 2D double-sided readout electrode equipped with discrete delay-lines and dedicated electronics. The THGEM electrodes, produced by standard printed-circuit board and mechanical drilling techniques, a 0.4 mm thick with 0.5 mm diameter holes spaced by 1 mm. Localization resolutions of about 0.7 mm (FWHM) were measured with soft x-rays, in a detector operated with atmospheric-pressure Ar/CH$_4$; good linearity and homogeneity were achieved. We describe the imaging-detector layout, the resistive-anode 2D readout system and the imaging properties. The THGEM has numerous potential applications that require large-area imaging detectors, with high-rate capability, single-electron sensitivity and moderate (sub-mm) localization resolution.

---

[*] Corresponding author.

# Contents

## 1. Introduction

Position-sensitive gaseous detectors are widely employed in particle physics and numerous other fields. With these, sub-millimetre localization properties in the detection of ionizing radiation, neutrons and soft x-rays are routinely achieved [1]. Moreover, following the very successful era of wire chambers, resistive-plate chambers and others, there have been many attempts to further improve the operational conditions and performance of gaseous detectors. Specifically, a variety of micro-pattern detectors, produced by state-of-the-art industrial techniques, has been proposed and is already in use in numerous radiation-detection fields [1, 2]. One of the most successful of these techniques is the Gas Electron Multiplier (GEM) [3], in which the multiplication occurs within tiny holes. Such hole-multiplication gives rise to confined avalanches and reduced secondary effects; high gains can thus be reached by cascading several GEM elements [4], resulting in single-electron detection sensitivities, as demonstrated in gaseous photomultipliers [5]. In addition to high gain, fast response and true pixelated radiation localization, GEMs have high counting-rate capability, superior to that of wire chambers [1].

    In this contest, the thick GEM-like gaseous electron multiplier (THGEM) [6-11], preceded by the "optimized GEM" [12] and "large electron multiplier" (LEM) [13], is one of the most recent developments in gaseous hole-multipliers. The THGEM is cost-effectively fabricated from double-clad G-10 plates, using standard printed-circuit board (PCB) techniques; it consists of mechanically drilled holes with chemically etched rims around each hole. The rim is crucial to significantly reduce the probability of gas breakdowns, resulting in high gains [8, 9]. THGEMs may be fabricated from a large variety of insulating materials. Due to their mechanical robustness, such devices may be produced to cover very large areas. The operation

mechanism and properties of the THGEM at atmospheric and at low gas pressures were described in detail in [8, 9]. An electric potential applied between the THGEM electrodes establishes a strong dipole electric field within the holes, giving rise to efficient focusing of ionization electrons into the holes and their multiplication in a gas avalanche process. The resulting electron avalanche might be collected on an anode or transferred to successive multiplier elements. The THGEM can operate in a large variety of gases; in some of these multiplication factors of $10^5$ with a single THGEM and $10^7$ with a cascaded double THGEM configuration, were achieved at atmospheric pressure [8]. The avalanche process is fast, with a typical rise-time of several nanoseconds, and the rate capability is in the range of MHz/mm$^2$ [8] at gains of $10^4$.

## 2. The THGEM imaging detector prototype

The double-THGEM imaging detector investigated in the present work is schematically shown in Fig. 1. It consists of a metallized-Mylar drift-cathode foil and two THGEMs coupled in cascade to a resistive anode; the latter was made of a > 2 M$\Omega$/square graphite lacquer layer sprayed on a G10 substrate. A double-sided strip readout electrode is placed behind the resistive-anode plate.

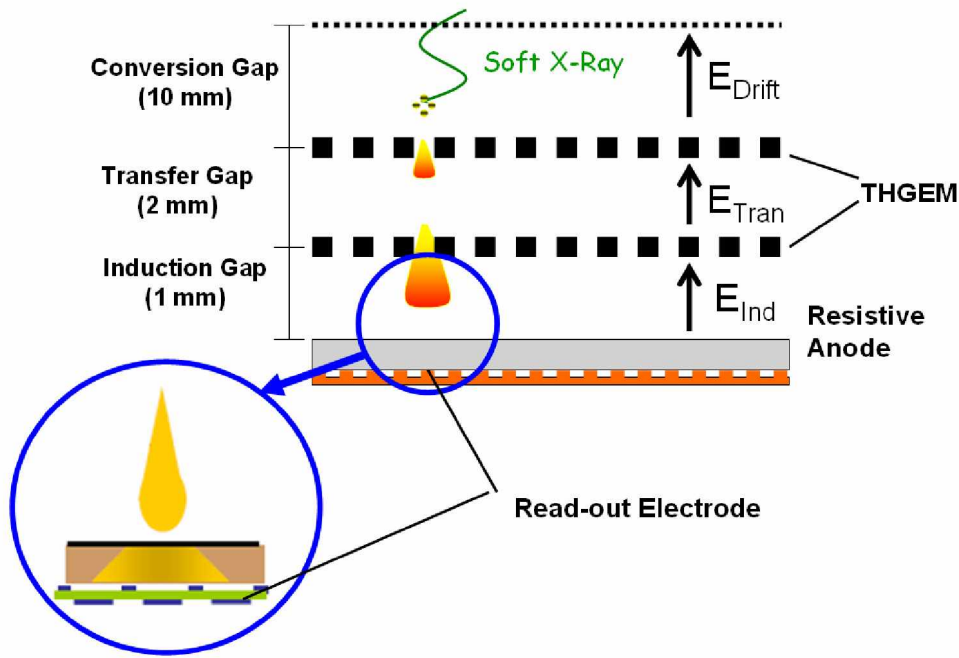

**Figure 1.** Schematic view of the position sensitive THGEM detector assembly. Radiation induced electrons are multiplied in a cascade within the THGEM holes; the resulting charge is collected onto a highly-resistive anode. The moving charge in the induction gap induces a localized signal at a position-encoding readout electrode.

Fig. 2 shows photographs of the THGEM and the readout electrodes. The THGEM electrodes employed in this work have holes of diameter 0.5 mm arranged in a hexagonal pattern, with a pitch of 1 mm; the etched rim around each hole is 0.1 mm. The thickness of the plate is 0.4 mm. The detector was located within a stainless steel vessel, 20 cm in diameter, that incorporates a 50 μm thick Mylar window. It was operated under continuous gas flow, at 1 atmosphere of Ar/CH$_4$, (95:5); the spacers between electrodes assured efficient gas exchange also in the amplification region. The radiation-conversion and drift gap above the first THGEM multiplier was 10 mm wide, while the transfer and induction gaps were 2 and 1 mm wide, respectively; the drift ($E_{Drift}$), transfer ($E_{Tran}$) and induction fields ($E_{Ind}$) were set at 1, 3.5 and 4.5 kV/cm, respectively. The position readout was realized by induced signal recording through the resistive anode onto a dedicated readout board, using delay-line encoding [14 - 19].

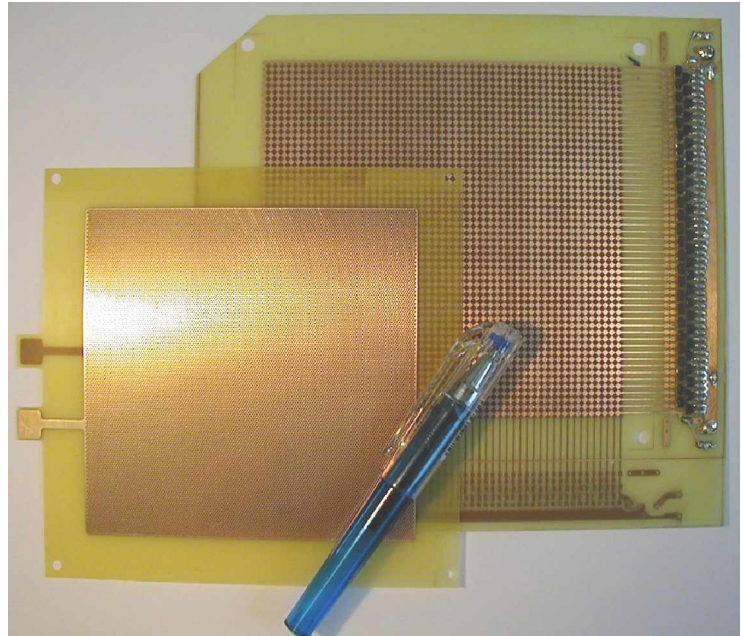

**Figure 2.** The THGEM and the double-sided readout electrode with pad-strips coupled to a discrete-element delay-line.

The resistive anode technique allowed for spreading of the induced signal on the readout electrode such that the geometrical size of the induced charge matches the width of the readout-strip pitch. This readout technique also permits for a galvanic decoupling between the multiplication stage and the readout electrode board. Thus, the resistive anode can be operated at high voltage while the readout board is maintained at ground potential. The low-capacitive coupling between resistive anode and readout board also protects the readout electronics from eventual spurious discharges in the detector [14].

The induced signals were collected on a double-sided X-Y readout electrode, structured with diamond-shape pads printed on both sides of a standard 0.5 mm thick printed-circuit-board (PCB) (Fig. 2). The pads are interconnected with strips running in orthogonal directions (X and Y) on each of the two board faces, with a pitch of 2 mm. The strips on each side of the board are coupled to a discrete delay-line circuit (Fig. 2) [14].

The pad electrode and the readout elements are shown schematically in figure 3. The printed diamond-shaped pads are geometrically designed such that equal charge is induced on both PCB sides and they are non-overlapping in order to reduce the capacitive coupling.

The avalanche location is derived from the time difference between induced signals propagating along the discrete LC delay-line circuit. The latter is composed of 52 LC cells (Fig. 3) with an inductance of L=290 nH and a capacitance of C=27 pF per cell; the corresponding delay is 1.4 ns/mm; the total delay for each coordinate is thus 140 ns and the nominal impedance is Z = 103 Ω.

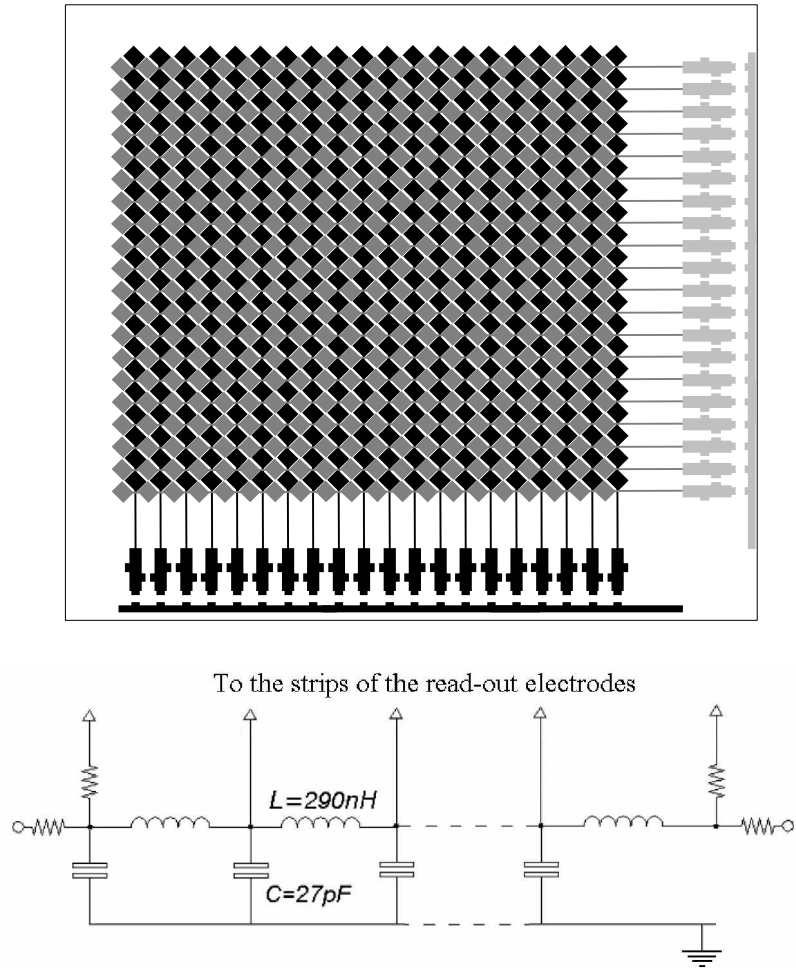

**Figure 3.** Schematic view of the readout electrode and the discrete-element LC delay-line circuit.

The detector was irradiated with 6 and 8keV x-rays from a $^{55}$Fe source, as well as an x-ray generator driven Cu-fluorescence source. Metal masks were placed in front of the detector for the imaging studies. In the localization-resolution measurements the distance between the detector and the x-ray source (1 mm in diameter) was 70 cm. The mask was positioned about 20 mm away from the drift cathode.

All THGEM electrodes were biased with independent high-voltage power supplies (CAEN N471A) through 20 MΩ serial resistors. The signals from the electrodes which are not at ground potential were capacitively decoupled from the high voltage. Signals from the electrodes of the second THGEM were fed to charge and current preamplifiers, providing the integrated charge signal and a fast common trigger for the position measurements and for coincidence and TOF experiments. For measurement of the total charge, the electron multiplication factor (gain) and the energy resolution, a charge-sensitive preamplifier (ORTEC 124 with a sensitivity of 275 mV/pC) and a shaping amplifier (ORTEC 572A) were used. A current-sensitive preamplifier (VV44, MPI Heidelberg) and a shaping amplifier (Ortec 570) were employed for the pulse-shape processing of the 'fast' component of the signals. The pulse-height spectra were analyzed by means of a multi-channel analyzer (Amptek MCA8000A).

## 3. The readout electrode and electronics

### 3.1 Optimization studies of the readout electrode

The position signals of the imaging THGEM detector were encoded with a pad-structured readout (R/O) electrode introduced by Eland et al [20] and Jagutzki et al. [21]. In previous works these electrodes were manufactured on thin Cu-clad Kapton foils. More rigid and economical R/O boards can be manufactured by the standard printed circuit (PCB) technique. In this case, the modified geometry required optimization of the readout strip/pad structures to ensure equal charge sharing and minimal cross-talk between both sides of the R/O electrodes.

a) *Optimization of the induced charge amplitudes on the electrode front and back sides*
The optimal geometry for the R/O electrode is the one with equal signal amplitudes induced on its front and back sides (i.e. signal amplitude ratio of front to back: $V_f/V_b = 1$). The front side denotes here the one closer to the resistive anode. The copper structure on the front side causes partial shielding of the board back side to the charge moving within the induction gap. Hence, the pads on the front side need to be smaller than those on the back side, to allow better signal transmission.

A systematic experimental study has been performed to optimize the R/O structure. Measurements were performed on small (35x35 mm$^2$) test boards, produced with different front-to-back pad- and strip-area ratios ($A_f/A_b$). Fig. 4 shows an example of front- and back-side patterns with an area ratio of 0.3:

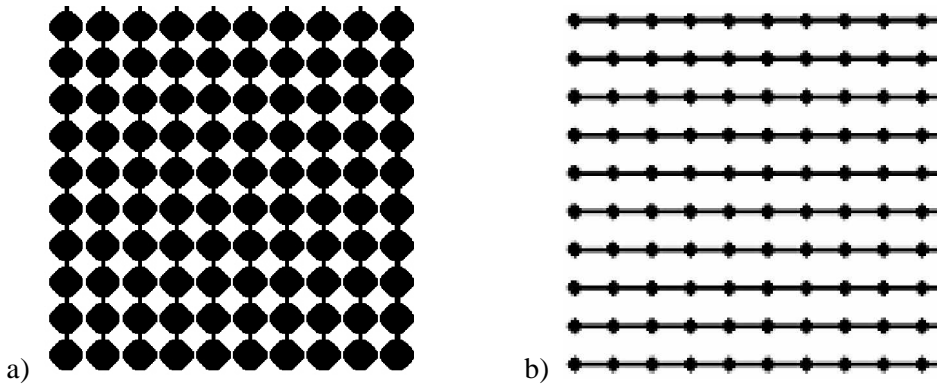

**Figure 4.** Example of back (a) and front (b) patterns of a readout test-board with a strip-and-pad area ratio $A_f/A_b = 0.3$.

The induced-charge measurements were performed with a small standard double GEM detector of 35x35 mm$^2$ size, irradiated by a collimated 5.9 keV $^{55}$Fe x-ray source. The experimental setup is schematically shown in Fig. 5.

Unlike the 2 MΩ/square carbon-sprayed resistive anode of the larger imaging THGEM detector, that of the small prototype was made by evaporating Ge onto a 1 mm thick float glass with a surface resistivity of the order of 30 MΩ/square. The choice of the resistive layer material is of no importance for these measurements and only a matter of convenience. The carbon lacquer technique was established as an economical method for producing appreciable quantities of large-area gas detectors [19]. The Ge-layers, on the other hand, developed for application in ultra-high vacuum devices, where high purity of the materials is mandatory [21]. As shown in Fig. 5, the distance (d) of the readout electrode to the Ge-layer could be varied. The induced signals on both sides of the readout electrode were recorded with a fast preamplifier; the amplitudes of the signals were analyzed with a digital oscilloscope (Tektronix TDS 3052A).

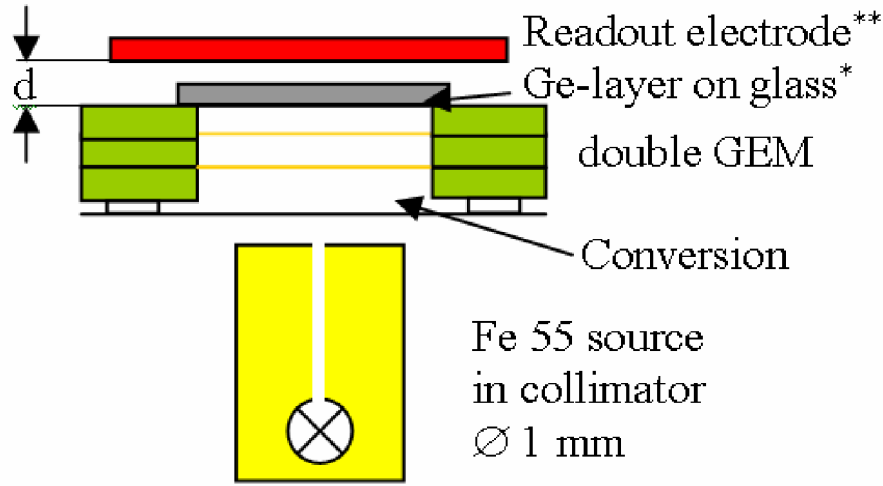

**Figure 5.** Measurement setup for optimizing the charge-amplitude distribution: a small test detector with two standard GEMs is irradiated with a collimated $^{55}$Fe source. A Ge resistive anode deposited on glass is coupled to a readout board, placed at a distance (**d**) from the Ge-film face.

Figs. 6a and 6b respectively show the two-sided readout board and an example of signals from its front ($V_f$) and back sides ($V_b$). The amplitude ratios ($V_f/V_b$) were obtained by averaging the measured amplitudes of 20 events for each R/O board, interchanging the linear amplifiers between both sides after each sequence of 10 measurements, to compensate for their different gains. The amplitude ratio was measured for a Ge-anode-to-R/O board distances of d = 1 mm.

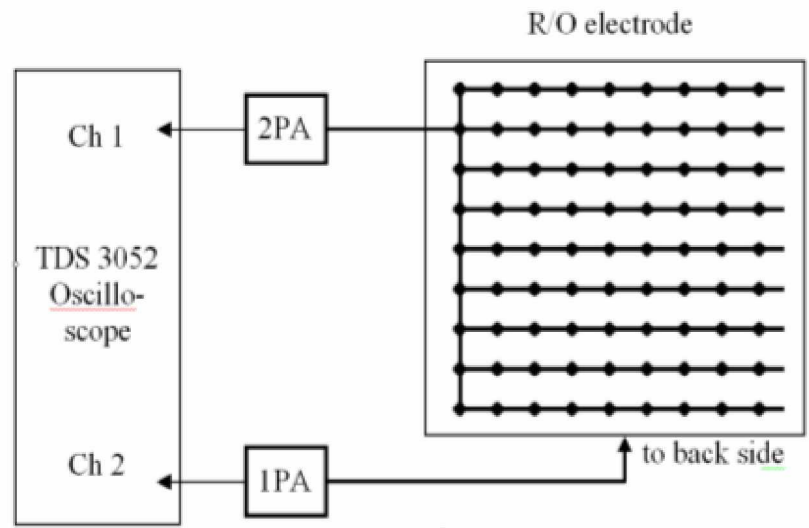

**Figure 6a.** Readout scheme for the amplitude ratios of the readout board

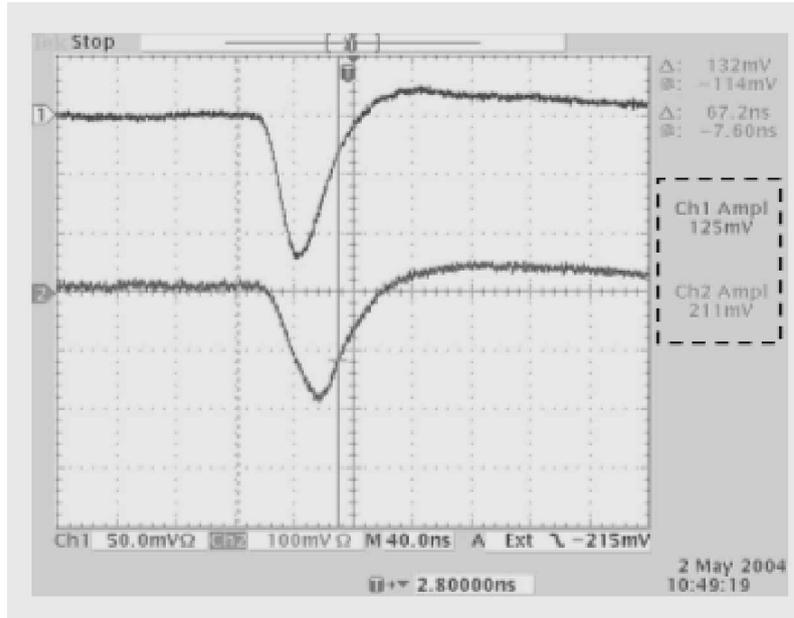

**Figure 6b.** Typical signals recorded from both faces of the board (see Fig. 6a). The amplitudes are indicated on the right side of the figure.

Fig. 7 shows the results for all measured R/O boards, for d = 1 mm (see Fig. 5). The optimal value ($V_f/V_b = 1$) at d = 1 mm was obtained at an area ratio $A_f/A_b$ = 0.27. It is not expected that the amplitude sharing will depend significantly on the distance between resistive anode and R/O board. This was confirmed by a measurement at d = 2.6 mm, in which equal amplitude sharing was obtained at the value $A_f/A_b = 0.27$ as well.

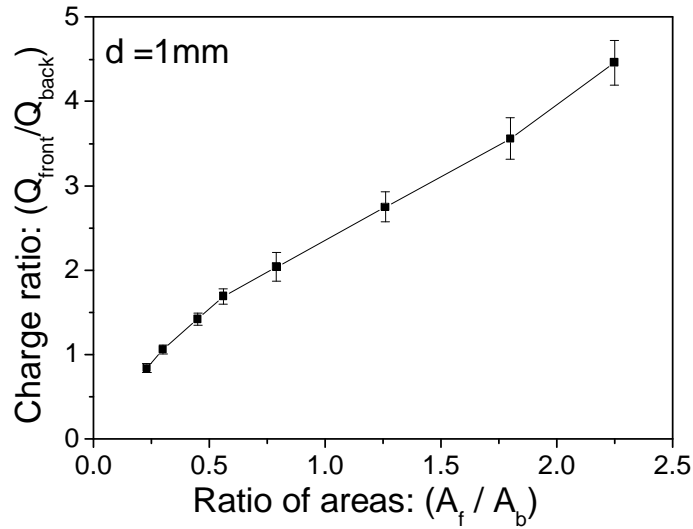

**Figure 7.** Results of measured charge-amplitude ratios, in the setup of Fig. 5, as function of the ratios of the areas covered by the pads and strips on the front ($A_f$) and back ($A_b$) sides of the investigated readout boards.

b) *Optimization of the radial spread of the induced charge*

An important factor governing the localization properties is the avalanche-induced radial charge spread. A too large spread leads to broadened signals at the output of the delay lines. This causes a reduction of the signal–to-noise ratio and a deterioration of the position resolution. On the other hand, a radial charge spread whose full width at half maximum (FWHM) is narrower than the pitch of the R/O strips causes modulation of the position response by the discrete structure of the R/O board. To obtain the highest position resolution without such modulation, the radial spread of the induced charge should be close to the pitch of the R/O strips [22]. Due to the highly-localized nature of the avalanche, this requirement is sometimes difficult to achieve in GEM detectors with direct charge readout. With the method of induced-charge readout via a resistive anode, the radial extension of the induced charge is principally determined by the distance between the resistive anode and the R/O board, and can thus be easily tailored to the needs of the R/O method.

For practical reasons (ease of manufacture, size of standard surface-mount devices (SMD) used for delay lines) we have chosen R/O boards with a pitch of 2 mm, requiring a similar FWHM radial charge spread. We have experimentally derived the radial charge distribution on the R/O board for two resistive film to board distances $d_1 = 1$ mm and $d_2 = 2.6$ mm. The experimental set-up is the same as used above for measuring the amplitude ratios (see Fig. 5). The readout method is shown in Fig. 8. The front-side of the R/O electrode was split; a central strip was connected to a fast linear preamplifier (PA1), the other strips being interconnected and fed to a second preamplifier (PA2). At the back side, all strips were interconnected into a third preamplifier (PA3).

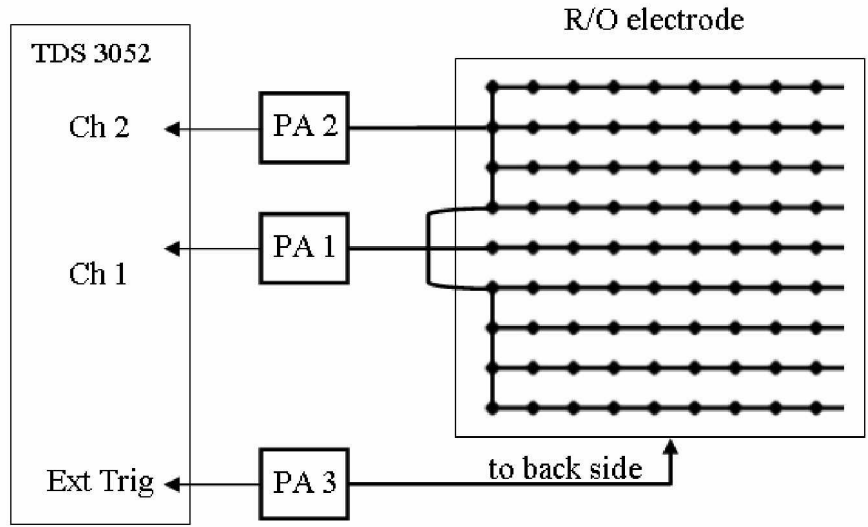

**Figure 8.** Schematic drawing of the readout circuit for measuring the radial charge distribution.

A well collimated $^{55}$Fe x-ray source was displaced, with about 1 μm precision, orthogonal to the strips; the amplitude of the fast current signals of PA1 – 3 were recorded in 500 μm steps with a fast digital oscilloscope (similar to Fig. 6b).

To reduce statistical fluctuations for each source position, the amplitude ratios of 12 events were measured for each location and averaged. Fig. 9a, 9b respectively show the amplitude ratios ($V_{PA1}/V_{PA3}$) of the central forward-side strip signal to that of the back side, as function of the x-ray source location, for distances d = 1, 2.6 mm.

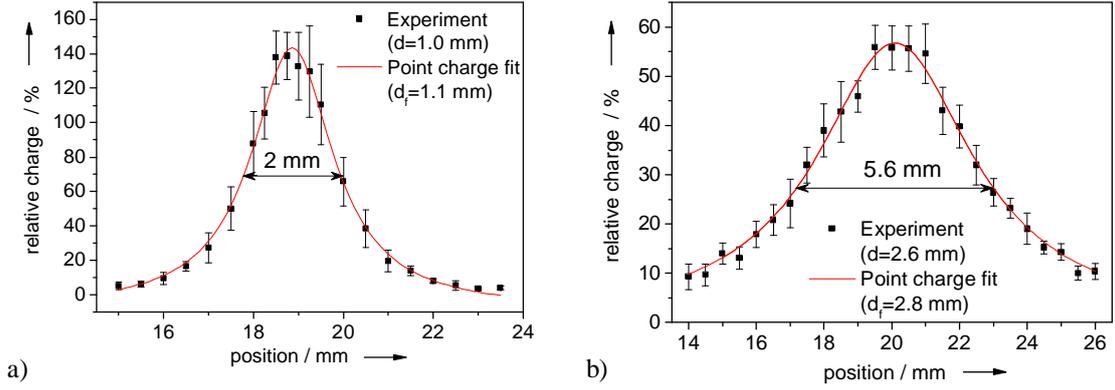

**Figure 9.** Projected radial distribution of the induced charge signal at a) 1 mm and b) 2.6 mm distance between resistive anode and RO electrode. The curves present fits of the experimental data using Lapington's model (see text).

Following Lapington's considerations [23], the mechanism for charge induction via capacitive coupling between the localized event charge on the resistive layer and the readout electrodes, causes the induced charge footprint to be spread in a very well defined manner. The resulting radial charge distribution is defined by:

$$\sigma_r \approx \frac{1}{(r^2 + a^2)^{3/2}} \qquad (1)$$

where $\sigma_r$ is the induced charge at horizontal distance $r$ from the point charge and $a$ is the distance between the point charge and the pick up electrode. The line in Fig. 9 presents a fit of the experimental data with the 1-dimensional projections of this distribution for 2 distances d. The distance $d_f$ in the fit was used as a free parameter and the resulting values for $d_f$ are in good accordance to the real distance in the experiment. The FWHM of the projected radial spread of the induced charge distribution is 2.0 and 5.6 mm for a distance between resistive anode and pick up electrode of 1.0 and 2.6 mm, respectively. For a R/O board with strips of 2 mm pitch the optimum spacing is therefore of the order d = 1 mm.

### 3.2 The readout electronics

Fig. 10 shows the scheme of the R/O electronics of the 100 x 100 mm$^2$ double-THGEM imaging detector. The data acquisition (DAQ) hardware is based on the 8-channel Time-to-Digital Converter (TDC) chip F1 and the ATMD board [24]. The system also comprises a Charge-to-Time Converter (QTC), based on LeCroy's MTQ300A-chip [25].

This module also permits measuring the avalanche charge behind the last THGEM (after pulse shaping with a Timing Filter Amplifier (TFA)). Events are stored in the PC memory with the corresponding position and time signals.

The signals from the ends of each delay-line and the "common start" signal from the last THGEM cathode were amplified by fast linear amplifiers (VV46, MPI-Heidelberg). The amplified position signals were delayed by 60 ns and fed to an ORTEC CF934 Constant Fraction Discriminator (CFD). The common signal was discriminated by a Canberra 1428 CFD and used as common start for the TDC. The slow (TTL) output of the CFD "enabled" the gate of the ORTEC CFD and allowed valid signals from the delay-lines to pass. The outputs of the CFD were used as "stop" signals for the TDC. The data acquisition software, a modified version of CAMDA [26], calculated the position coordinates and performed a plausibility check on the measured timing signals (comparison of time-sums). Valid data were accumulated in histograms and/or stored as list-mode files in the PC memory.

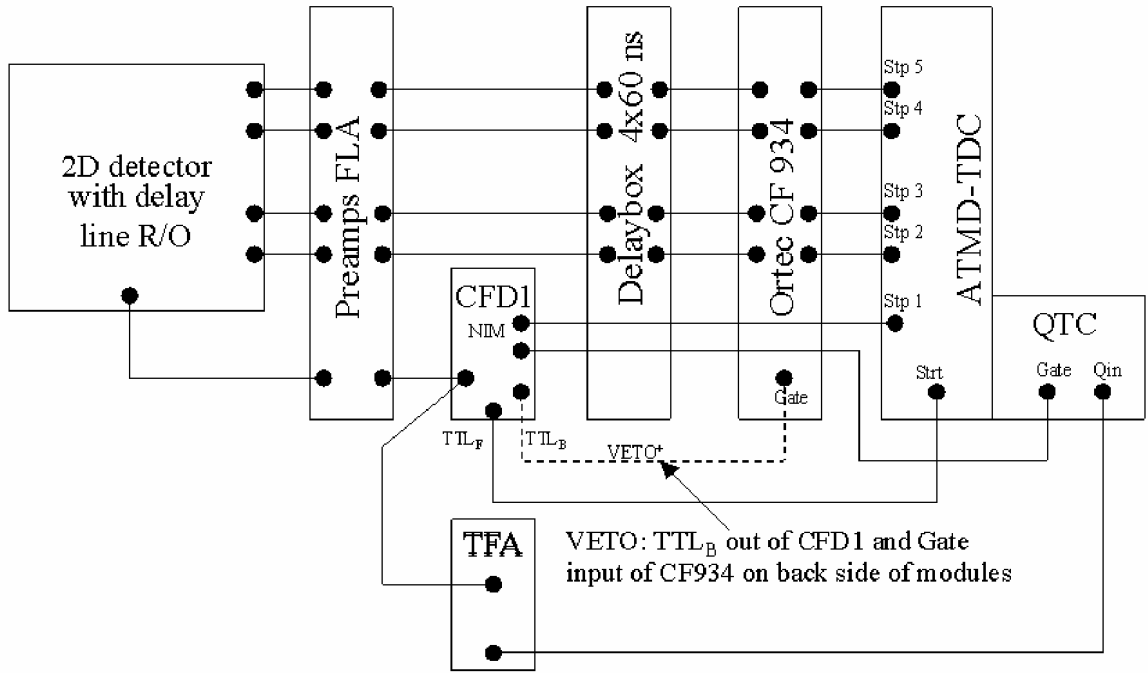

**Figure 10.** Schematic view of the DAQ system.

The result of the digitalization process is an image of 800x800 pixels; each pixel has a linear extension of 125 μm. The images are stored with high dynamic range (4 Byte integers / pixel). However, for visualization of the image on the screen, these high-dynamic-range images were compressed to 8-bit gray-level images.

## 4. Results

The performance of the THGEM imaging detector was characterized through the analysis of the charge signal, namely gain uniformity and energy resolution, and the measurements of some parameters as the integral non-linearity (INL), the homogeneity and the digital noise. The characterization of the imaging detector response in terms of spatial resolution, or more precisely in terms of spatial frequency response, was computed by the determination of the point spread function (PSF) and the contrast transfer function (CTF). From CTF the limiting spatial resolution was calculated.

### 4.1 Gain uniformity and energy resolution

The gain uniformity of the cascaded THGEM detector was investigated with a collimated 5.9 keV $^{55}$Fe x-ray source, 1 mm$^2$ in diameter. The active area was irradiated in 15 mm steps in both directions, for a total of 25 measurement points across the detector. The charge signal was derived from the cathode of the second THGEM. Fig. 11 shows examples of gain curves in Ar/CH$_4$ (95:5) and in Ar/CO$_2$ (70:30) at atmospheric pressure, obtained at the center of the active area.

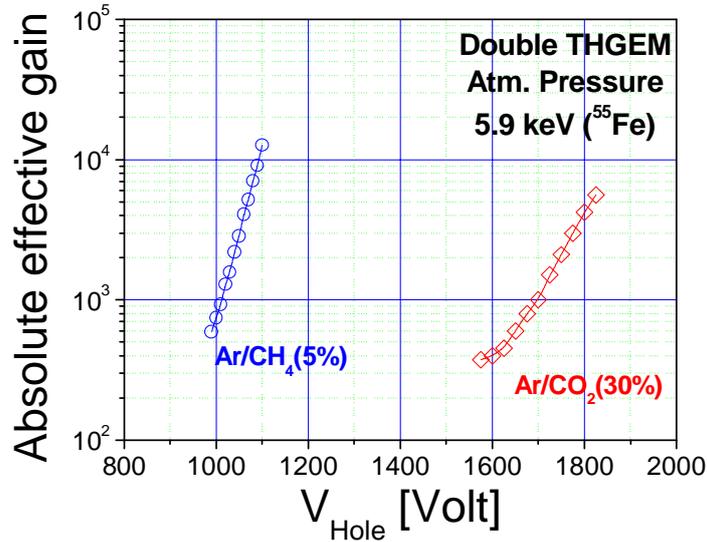

**Figure 11.** Absolute effective gain obtained at the center of the active area of 10x10 double-THGEM multipliers, measured with 5.9 keV x-ray ($^{55}$Fe) in Ar/CH$_4$ (95:5) and in Ar/CO$_2$ (70:30), at atmospheric pressure.

As shown in Fig. 12a, for a potential difference (V$_{hole}$) of 1220 V applied to both THGEMs, the absolute effective gain is rather uniform with an average value of 7x10$^3$; for 5.9 keV photons this corresponds to an average avalanche size of the order of 2x10$^6$ electrons. The overall measured gain variation is 10% FWHM. A typical energy spectrum of the $^{55}$Fe source is shown in Fig. 12b; the energy resolution of the photo-peak is of the order of 21% (FWHM), typical for gas avalanche detectors.

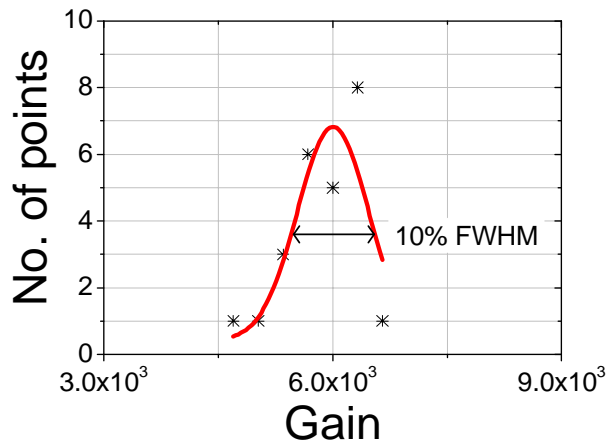

**Figure 12a.** Gain distribution of a collimated 5.9 keV $^{55}$Fe source over the sensitive area of the cascaded THGEM detector, measured at 25 points across its area in Ar/5%CH$_4$; V$_{hole}$ = 1220 V for each THGEM.

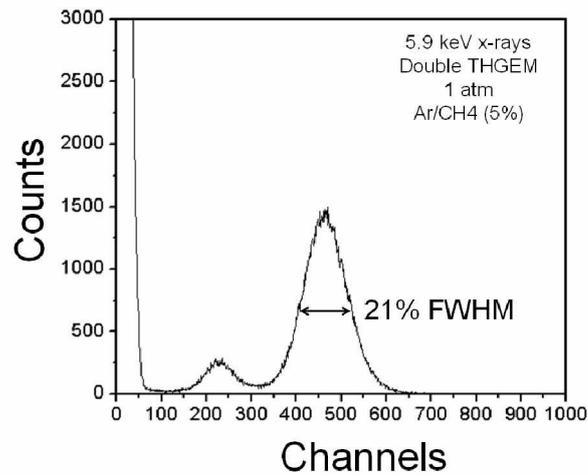

**Figure 12b.** Typical pulse-height distribution of the collimated source; the energy resolution is 21% (FWHM).

## 4.2 Imaging linearity and homogeneity

The image linearity and the homogeneity of the imaging system response were assessed by investigating the Gaussian fit of some pinholes' image projections (Fig. 13b). The pinholes were of 1 mm diameter, drilled into a 1 mm thick brass mask (Fig 13a). The detector was uniformly irradiated by characteristic K-shell fluorescence photons from a Cu-target (8 keV), excited by a well collimated intense bremsstrahlung-spectrum from an X-ray generator with a Mo anode.

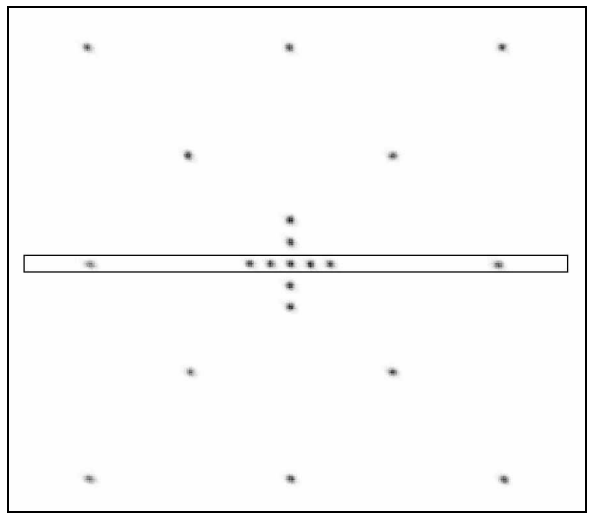

**Figure 13a.** X-ray transmission image of a mask with 1 mm diameter pinholes, recorded with the THGEM imaging detector. The projection profile of the selected area is shown in Fig. 13b.

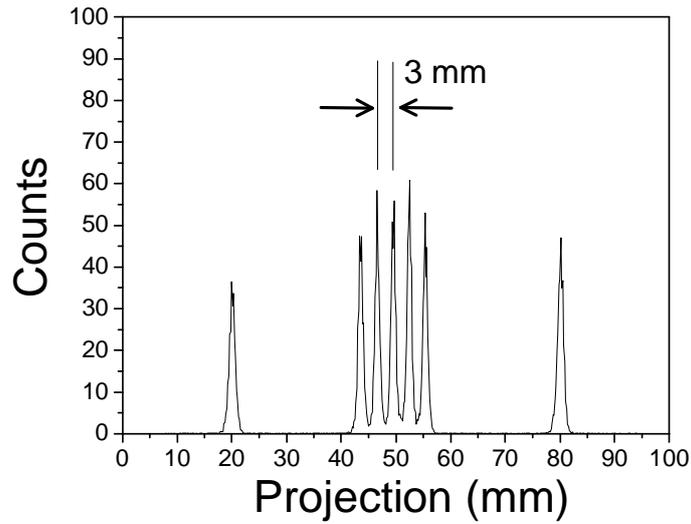

**Figure 13b.** Raw data projections of some of the pinhole images marked in Fig. 13a.

The effects induced on the image by distortions can be characterized by the integral non-linearity (INL). INL is defined as the deviation of the measured holes' centroid from their real locations. Geometrical distortions represent a deviation from rectilinear projection (Fig. 13c). They may arise from several effects, such as defects on the grid cathode or signal deformation as they propagate along the delay line. Specifically, alteration of the signal shape and its rise time depend on the amplitude of the signal (SNR), preamplifier cross talk, capacitive cross talk between read-out electrodes and asymmetry of induced signal on front and back. We measured an average INL value lower than 0.1% over the full image range of 100 mm.

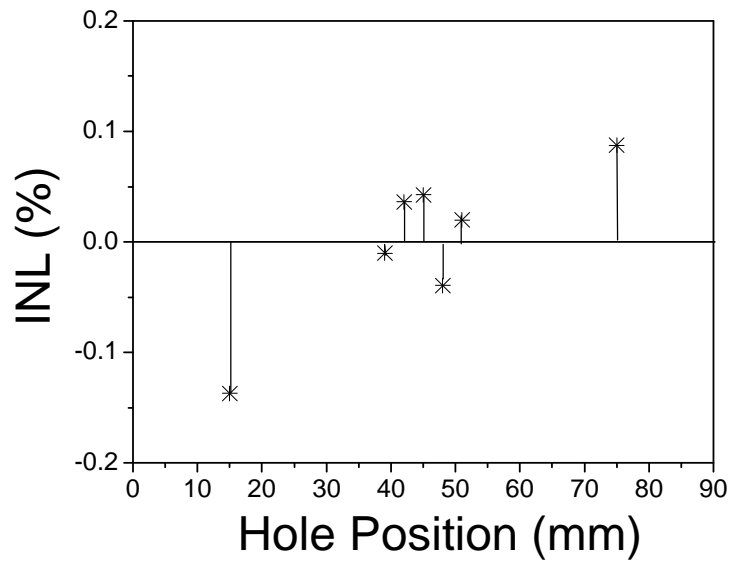

**Figure 13c.** Computation of the INL (integral nonlinearity) along the selected image projection.

## 4.3 Spatial resolution and CTF

The point spread function (PSF) of the imaging detector was obtained by means of the edge spread function (ESF) technique [27]. The ESF was measured by irradiating the edge of a 1 mm thick Aluminium plate (see 2D image in Fig. 14a) and extracting a projection of the edge profile from the image data (Fig. 14b).

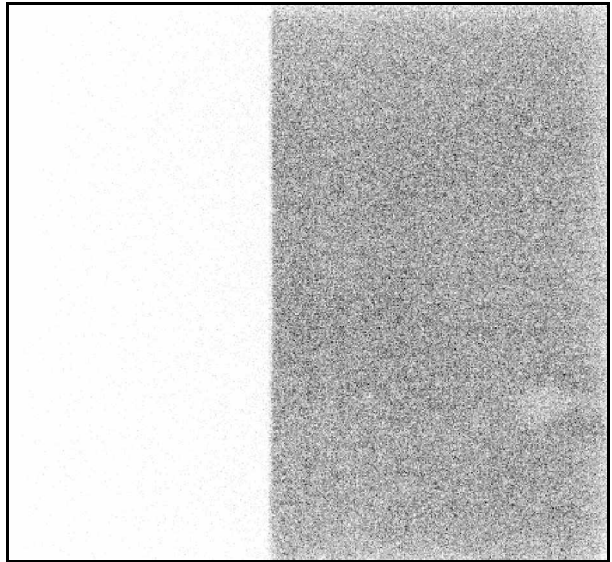

**Figure 14a.** A 2D image of a knife-edge obtained with the THGEM imaging detector, irradiated with a broad Cu X-ray beam.

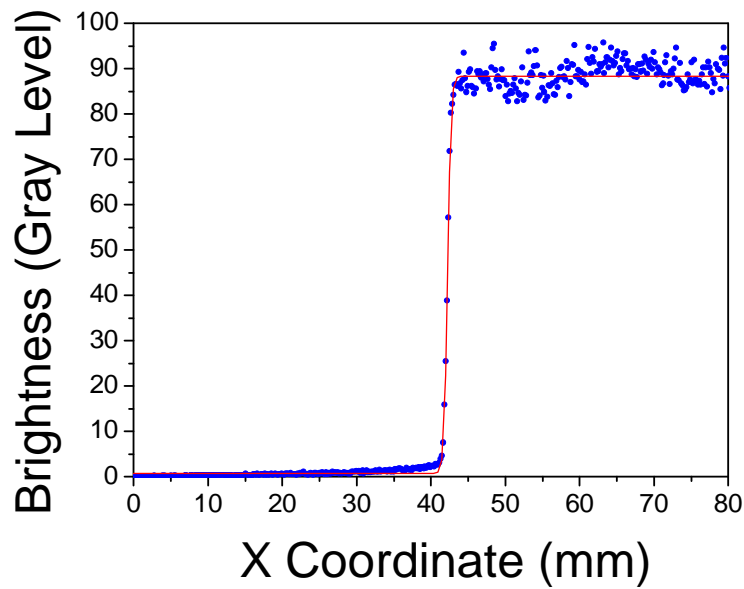

**Figure 14b.** The edge spread function along the x coordinate (blue dots). The red curve is a fit with a suitable model function.

In order to eliminate fluctuations resulting from the digital noise, we fitted the edge profile with an empirical function which models the ESF with adequate accuracy. The empirical model function used is given by the following equation [28]:

$$\text{ESF}(x) = a_0 + \frac{a_1}{1 + \exp(-a_2(x - a_3))} \qquad (2)$$

where $a_0$ is related to the transmission of the Al plate, $a_1$ is the brightness of the full irradiated sensitive area, $a_2$ is the steepness of the edge function (related to the spatial resolution of the imaging system) and finally $a_3$ is the centroid of the edge function (see Fig. 14b).

The determination of the PSF was obtained in a straightforward way by simply differentiating the empirical model of the fitted ESF. Consequentially, the derived FWHM of the resulting PSF, which is a direct manifestation of the intrinsic spatial resolution of the system, is then given by:

$$\text{FWHM} = \frac{3.53}{a_2} \qquad (3)$$

To check the consistency of our estimate of the spatial resolution, the PSF and its FWHM were determined by two different modalities (Fig. 15): from the numerical differentiation of the measured ESF (blue line), and the differentiation of the suggested empirical model (red line). There is good agreement between the measured PSF and the one obtained from the empirical model calculations. Nevertheless the direct analysis of the raw data, fitted with a Gaussian function distribution, has large fluctuations due to the digital noise (Fig. 15). Therefore, the related Gaussian fitting and the resultant spatial resolution has a larger uncertainty.

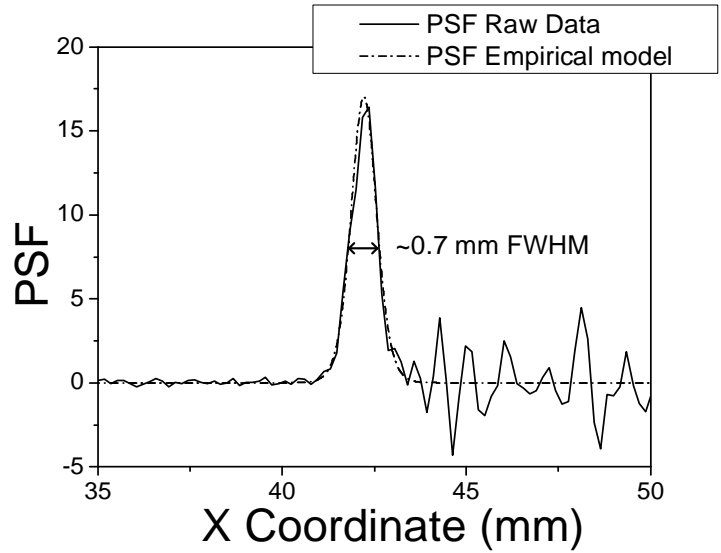

**Figure 15.** Results of the PSF determination along the x-coordinate using the differentiation of raw data (solid line) and the differentiation of the empirical model results (dashed line)**.**

The spatial resolution derived via the "measured" PSF, from raw data analysis, has a FWHM value of 0.71±0.05 mm, while the empirical model analysis reveal a FWHM value of 0.67±0.02 mm.

The result of imaging masks with a rectangular transmission profile at different fundamental frequencies is a frequency-dependent variation of the modulation depth (Contrast Transmission) in the images [29]. The CTF approaches a value of 100% at very low spatial frequencies, corresponding to a wide spacing period, and gradually drops with increasing spatial frequency. In general, 100% contrast represents an image of regular periodic structure of white and black lines, while 0% contrast is manifested by gray bars that merge into a gray background of the same intensity. When the contrast value reaches zero, the image becomes uniformly gray and remains so for all higher spatial frequencies.

The specimen modulation (contrast), $M(\nu)$, for patterns at different frequencies, can be defined as:

$$M(\nu) = \frac{I(max) - I(min)}{I(max) + I(min)} \qquad (4)$$

where $\nu$ is the spatial frequency of the grating specimen, I(max) is the maximum intensity transmitted by a repeating structure (grating) and I(min) is its minimum intensity. The contrast transfer function (CTF) is defined as the modulation depth of the image ($M_i$) divided by the modulation depth of the stimulus ($M_o$), as shown in the following equation:

$$CTF(\nu) = \frac{M_i(\nu)}{M_o(\nu)} \qquad (5)$$

The CTF is normalized to unity at the lowest spatial frequency, where the imaging system transfers the maximal contrast.

We have irradiated a brass mask placed in front of the THGEM detector, composed of different frequency periodic line gratings, using the Cu-fluorescence photon beam. The 2D mask image is shown in Fig. 16a; the projected distributions of the mask image are shown in Fig. 16b. The calculated CTF is shown in Fig. 17. The limiting spatial resolution, defined as the frequency at which the CTF drops to 10% of its maximal value, is of the order of 1.25 lp/mm, corresponding to a wave length of 0.8 mm.

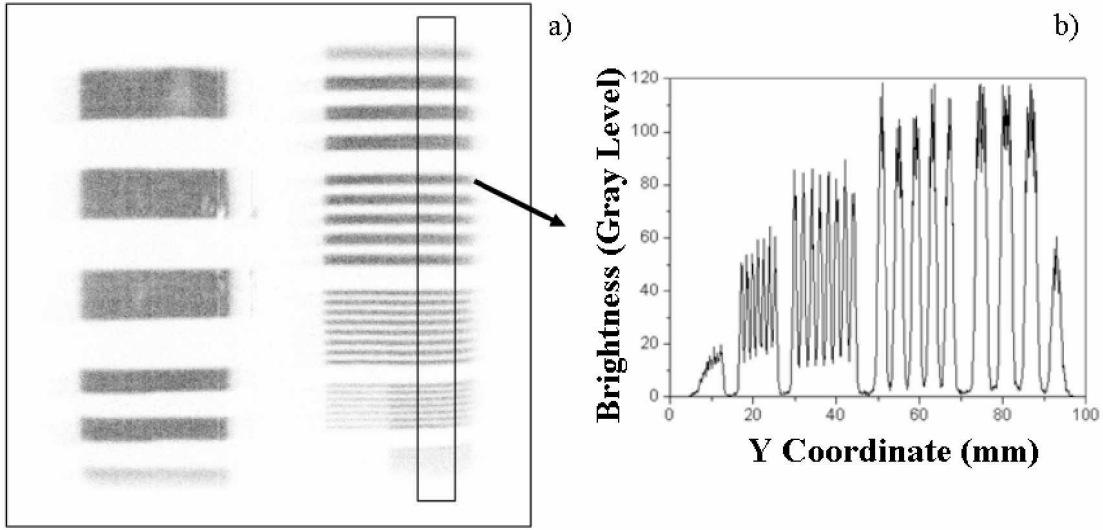

**Figure 16.** The imaging detector X-ray image of a 2D mask, used for the CTF evaluation (a) and the related projection of the selected area (b).

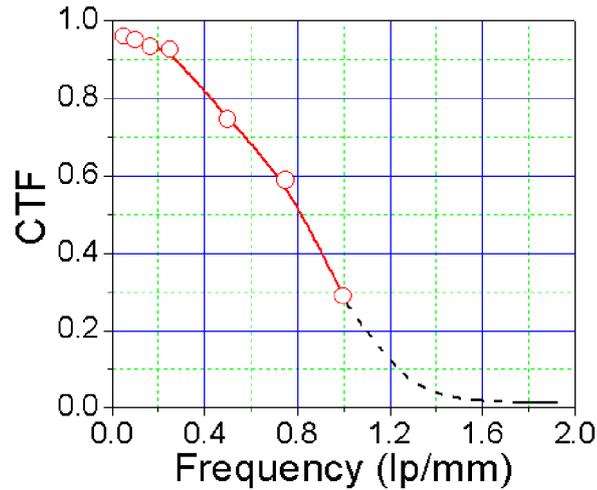

**Figure 17.** The contrast transfer function obtained from the mask image of Fig.16.

The image distortions at the extremities of the sensitive area of the detector, resulting in a slight increase of the INL (see Fig 13c), are due to the rather close irradiation geometry. Notice that the source-to-mask distance was only 70 cm in our experiments; the mask had a step-like profile with 2 cm thickness in the center and 1 cm at the edge, causing a partial shadowing effect on the edge of the detector's sensitive area. Finally, the optical magnification, defined as the ratio between the apparent size of an object (in our case the pinhole) and its true size, is another factor which might have influenced the performance of the imaging system. The linear-optical magnification, in the case of a point-like source, depends only on the source-object distance and object-conversion gap; in our case the linear optical magnification effect, estimated to be of the order of 1.036, is rather small.

**4.4 Digital noise**

Electronic imaging, like all imaging techniques, is always affected by blur and noise. The noise essentially arises due to processes which precede the production, capture, conversion and interpretation of the real source signal. The quantification of the noise is crucial for the analysis and optimization of an imaging systems' performance. For this reason we have determined the intrinsic signal-to-noise ratio of our imaging system the analysis of a flat-field image (Fig. 18a), obtained while irradiating the detector with a homogeneous ("flat") 8 keV photon beam. The image was recorded for 300 minutes of irradiation at intensities around 1 kHz/mm$^2$. The image appears to be homogeneous across the entire sensitive area, which also reflects the gain homogeneity previously discussed; the small slab in the lower part of the image is due to a defect of the resistive anode.

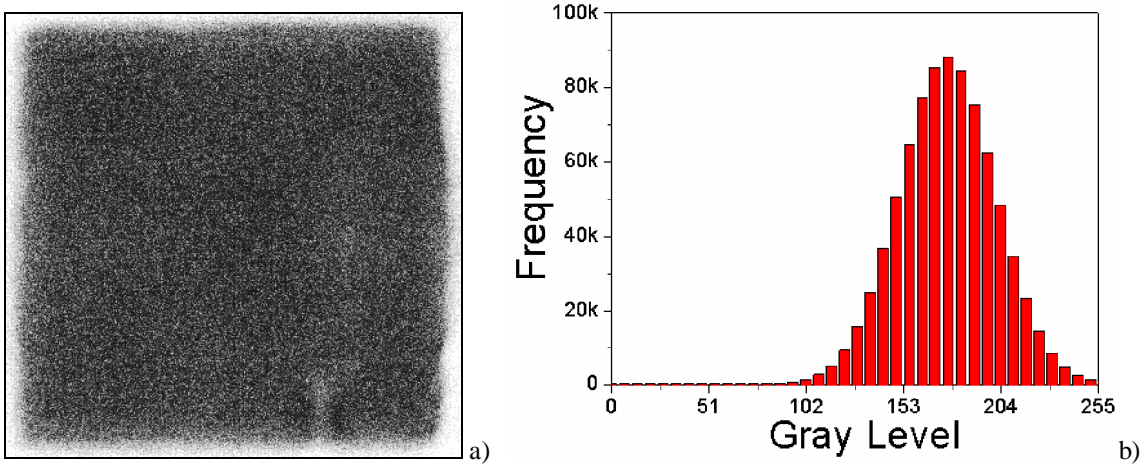

**Figure 18.** Flat-field image recorded by homogenous detector irradiation (a) and the histogram of the gray-level distribution of the flat-field image (b). See text.

Fig 18b shows the distribution of the flat image pixel contents. The average pixel content is A = 175 (gray level) with a standard deviation of σ = 25, corresponding to an average spread of about 14% (1 std. dev.). In terms of signal-to-noise ratio:

$$\text{SNR} = 20 \log_{10}\left(\frac{A}{\sigma}\right) = 16.9 \text{ dB} \qquad (6)$$

The resulting digital noise is not due to counting statistics (which, by itself, would result in an average spread of only σ = 5% per pixel). It is rather due to nonlinearities in the delay line readout (variation of the L's and C's of the delay line, see Fig. 3), cross talk of the readout channels and nonlinearities of the TDC. The digital SNR appears to be uniform across the sensitive detector area, with a slight degradation towards the image edge. This effect is essentially due to distortions of the electric fields near the edge of the THGEM electrodes. Also, position sensitive gain variations in the detector lead to degradation of the general performance of the imaging system, however, image processing can largely correct these defects and restore image uniformity almost to the quantum noise limit.

## 5. Summary and discussion

In view of the many possible potential applications of THGEM-based imaging detectors, we have extensively studied the localization properties and quality of imaging a 2D 10x10 cm$^2$ detector prototype, that operates in Ar/5%CH$_4$ at atmospheric pressure and room temperature.

An energy resolution of 21% FWHM was measured for 5.9 keV X-rays with a detector gain of 6x10$^3$. The gain uniformity was within ~10% over the detector's 10x10 cm$^2$ sensitive area. The performance of the imaging system was determined by invoking the common transfer theory tools (PSF and CTF), and digital SNR. For low energy x-rays the average intrinsic width of the PSF over the entire sensitive area is about 0.7 mm (FWHM), using a THGEM electrode with 0.5 mm hole diameter, 1 mm hole spacing and a readout electrode with a pitch of 2 mm.

The development of large area position-sensitive detectors is motivated in particle physics by the quest for rare events, to be discriminated against an extremely high background [1]. In the case of detecting charged particles, a very high position resolution could be achieved, as, for example, in tracking measurements: ~ 12 μm localization resolution was achieved with micromegas [30] and ~ 40 μm with the GEM [31]. Although the high granularity of micro-

– 18 –

pattern detectors plays an important role in these applications, offering excellent position resolution properties, there still remains a serious problem of discharges which cause critical damage to electrodes. Conversely, THGEM-based detectors may become an attractive alternative solution for numerous applications, where large-area radiation imaging detectors with modest (sub-mm) position resolution [9] are required. They are robust-element, self-supporting, and can be economically produced with simple printed-board techniques. They are capable of reaching single- or few-electron detection sensitivities, even for fluxes approaching several MHz mm$^2$ [9]. Potential applications are: UV photon detectors (e.g. with CsI photocathode) [7], particle tracking, (e.g. large-area muon trackers), sampling elements in calorimetry, x-ray imaging etc. THGEM multipliers coupled to solid radiation converters can be applied to neutron detection [14]. Feasibility studies of THGEM elements in cryogenic Dark Matter detectors are in progress, using low-radioactivity material substrates [32].

*Matter Detector,* to be published in the Proceedings of the 23rd Winter Workshop on Nuclear Dynamics, Big Sky, Montana (2007).